# A new role for exhaled nitric oxide as a functional marker of peripheral airway caliber changes: a theoretical study.

**Karamaoun C[1], Haut B[1], Van Muylem A[2]**

______________________________________________________________

[1]Ecole polytechnique de Bruxelles, Transfers, Interfaces and Processes, Université libre de Bruxelles, Brussels, Belgium

[2]Chest Department, Erasme University Hospital, Université libre de Bruxelles, Brussels, Belgium

**Address for correspondence**:
Alain Van Muylem,
Chest Department
Erasme University Hospital, Université libre de Bruxelles
808 Route de Lennik
1070 Brussels - Belgium
Phone: +32-2-555-39-53
Fax: +32-2-555-44-11

E-mail: avmuylem@ulb.ac.be

**Running head**: Exhaled nitric oxide as an airway caliber change marker

**Key words**: Modeling, exhaled nitric oxide, airway caliber, functional index.

**Funding**: Study funded by a MAP project from ESA: Airway NO in space

**ABSTRACT**

Though considered as an inflammation marker, exhaled nitric oxide ($F_E$NO) was shown to be sensitive to airway caliber changes to such an extent that it might be considered as a marker of them. It is thus important to understand how these changes and their localization mechanically affect the total NO flux penetrating the airway lumen (JawNO), hence $F_E$NO, independently from any inflammatory status change.

A new model was used which simulates NO production, consumption and diffusion inside the airway epithelium wall, then, NO excretion through the epithelial wall into the airway lumen and, finally, its axial transport by diffusion and convection in the airway lumen. This model may also consider the presence of a mucus layer coating the epithelial wall.

Simulations were performed that showed the great sensitivity of JawNO to peripheral airways caliber changes. Moreover, $F_E$NO showed distinct behaviors depending on the location of the caliber change. Considering a bronchodilation, absence of $F_E$NO change was associated with dilation of central airways, $F_E$NO increase with dilation up to pre-acinar small airways, and $F_E$NO decrease with intra-acinar dilation due to amplification of the back-diffusion flux. The presence of a mucus layer was also shown to play a significant role in $F_E$NO changes.

Altogether, the present work provides theoretical evidences that specific $F_E$NO changes in acute situations are linked to specifically located airway caliber changes in the lung periphery. This opens the way for a new role for $F_E$NO as a functional marker of peripheral airway caliber change.

**NEW AND NOTEWORTHY**

Using a new model of NO production and transport allowing realistic simulation of airways caliber change, the present work provides theoretical evidences that specific exhaled nitric oxide ($F_E$NO) changes, occurring without any change in inflammatory status, are linked to specifically located airway caliber changes in the lung periphery. This opens the way for a new role for $F_E$NO as a functional marker of peripheral airway caliber change.

# INTRODUCTION

Exhaled nitric oxide (NO) essentially results from NO produced in the airway epithelium diffusing into the airway lumen during expiration (27). Since NO epithelial production has been shown to be triggered by TH2 type airway inflammation (4), it was mainly regarded as a non-invasive inflammation monitoring tool in asthma (8).

However, besides inflammation, acute changes in airway caliber are also a primary feature of the asthma pathology and, for more than a decade, several works showed that reductions of airway caliber, induced by airways challenges, lead to a dramatic decrease of exhaled nitric oxide concentration ($F_E$NO) (7, 13, 19, 21). This decrease is even able to mask the impact on $F_E$NO of an ongoing inflammatory process (12).

Moreover, different patterns of $F_E$NO response to β2-agonists were recently demonstrated in asthma patients (20). These distinct behaviors, *i.e.* stability, significant increase and decrease, were related to specific behaviors of ventilation distribution indices, reflecting airway caliber changes at different depths of the bronchial tree. So, besides its role as an inflammatory marker, $F_E$NO is potentially a functional marker of amplitude and location of airway caliber changes.

This emphasizes the importance of understanding how the caliber and location of an airway affects the amount of NO penetrating the lumen from the surrounding epithelium. One work (32) partly tackled this issue using a model of NO transport incorporating convection and molecular diffusion acting in realistic boundary conditions. However, though informative, this model was not designed to simulate the passage of NO from the epithelium to the airway lumen and the way lumen caliber changes affect it.

Therefore, a new model was developed which realistically describes the airway tissue layers: smooth muscle, epithelium, and mucus; and the airway caliber changes and which may simulate NO transport in the airway lumen as well as in the epithelium wall. This model was presented elsewhere in its technical and mathematical aspects (16). The aim of the present work is to theoretically assess whether $F_E$NO changes after acute manoeuvers reflect amplitude and specific location of airway caliber changes and, thus, whether $F_E$NO, besides its role as an inflammatory marker, may also be a functional marker.

## METHODS

The detailed mathematical features of the model used in this paper are extensively described in Karamaoun *et al.* (16). This model integrates NO production and transport inside the epithelium, the muscle and the mucus layers, the excretion from epithelium inside the airway lumen and the axial transport of NO in the lumen, the latter feature being derived from Van Muylem *et al.* (31). The geometrical boundaries are derived from Weibel's symmetrically dichotomic model of the lung (33), composed of 24 generations (generation 0 being the trachea) and associating one length and one airway diameter to each generation.

a) <u>NO production, exchange and transport in the bronchial tissues</u>

From generation 0 to 18, the airways are assumed to be surrounded, from outer side to lumen side, by a blood layer, a double tissue layer representing the muscle and the epithelium, and a mucus layer as described on Figure 1 (Panel C).

NO production occurs in the epithelial layer (11), at a constant volumetric rate *Pr*.

Once produced, a NO molecule may:

1) diffuse and be consumed in the epithelium or in the muscle layer,
2) diffuse and be absorbed by the blood layer, which may be considered as an infinite sink due to the high affinity of NO for blood hemoglobin (6),
3) diffuse through the mucus layer and be excreted inside the airway lumen.

In the $i^{th}$ generation (subscript *i*), 4 equations describe the transport of NO in the blood (Eq.1), the muscle (Eq.2), the epithelium (Eq.3) and the mucus (Eq.4).

$$C_{t,i} = 0 \quad \text{for} \quad x = 0 \quad (1)$$

$$D_{NO,t} \frac{d^2 C_{t,i}}{dx^2} - k\, C_{t,i} \quad \text{for} \quad 0 < x < \delta_M \quad (2)$$

$$D_{NO,t} \frac{d^2 C_{t,i}}{dx^2} + Pr - k\, C_{t,i} = 0 \quad \text{for} \quad \delta_M < x < \delta_M + \delta_E \quad (3)$$

$$D_{NO,t} \frac{d^2 C_{t,i}}{dx^2} = 0 \quad \text{for} \quad \delta_M + \delta_E < x < \delta_M + \delta_E + \delta_\mu \quad (4)$$

As depicted Fig.1C, the *x* axis is a cartesian coordinate normal to the inner airway surface and originating at the muscle/blood interface; $\delta_M$, $\delta_E$ and $\delta_\mu$ are the muscle, epithelium and mucus thicknesses, respectively. $C_t$, is the NO concentrations in the tissue, $D_{NO,t}$ is the diffusion coefficient of NO in the tissues and mucus assimilated to water, $k$ is the kinetic constant of the reaction consuming NO in the epithelium and muscle tissues. Indeed, this reaction was shown to mainly happen with superoxide and metalloproteins and to be of the first order with respect to NO concentration (3).

These equations are written using a quasi-steady state approximation, in cartesian coordinates. These two assumptions are extensively discussed in Karamaoun *et al.* (16).

b) <u>Tissue-air interface</u>

In the airways, the tissue-air interface corresponds to $x = \delta_M + \delta_E + \delta_\mu$. Assuming a thermodynamic equilibrium between the NO concentration dissolved in the tissue ($C_t$) and the lumen gaseous ($C$) NO concentration, the following equation can be written, for the $i^{th}$ generation,

$$C_{t,i}(x = \delta_{M,i} + \delta_{E,i} + \delta_{\mu,i}) = \lambda_{t:air} \, C_i \quad (5)$$

The equilibrium constant $\lambda_{t:air}$, is based on the Henry's constant for NO in water.

The expression of the NO flux density between the bronchial wall and the gas phase, expressed in m.s$^{-1}$, can thus be established, based on the NO concentration at the interface (29), as

$$J_{air,i} = -\gamma \, D_{NO,t} \frac{dC_{t,i}}{dx} \big|(x = \delta_{\mu,i} + \delta_{E,i} + \delta_{M,i}) \quad (6)$$

where $\gamma$ is a dimension factor allowing to express $J_{air}$ in m.s$^{-1}$ (16). An airway epithelial production ($J_{air}$) is considered up to generation 18. Beyond, the NO production is assumed to come from the alveolar epithelium.

In the alveoli, for the $i^{th}$ generation, the NO exchange flux density between the alveolar wall and the gas phase, expressed in m.s$^{-1}$, can be written, following Van Muylem *et al.* (31), as

$$J_{alv} = (P_{alv} - U_{alv}C)/S_{alv,tot} \quad (7)$$

where $P_{alv}$ is the total alveolar production rate and $U_{alv}$ is the NO consumption in the alveolar tissues. $S_{alv,tot}$ is the total lateral surface of the alveoli, calculated from the Weibel's data (33), assuming an alveolar diameter of 0.2 mm.

c) <u>Axial NO transport in airway lumen</u>

Eq. 8 describes the axial diffusion-convection transport for NO in the $i^{th}$ generation.

$$\frac{\partial C_i}{\partial t} = -\frac{Q_i}{\Omega_i}\frac{\partial C_i}{\partial z} + D_{NO,air}\frac{\Omega'_i}{\Omega_i}\frac{\partial^2 C_i}{\partial z^2} + \frac{1}{L_i \Omega_i}(J_{alv,i}\, S_{alv,i} + J_{air,i}\, S_{air,i}) \qquad (8)$$

where $z$ is the axial coordinate originating at the alveolar end and $t$ is the time. $Q$ is the air flow; $\Omega$, $\Omega'$ and $L$ are the total cross-sectional area (bronchi + alveoli), the airway cross-sectional area and the length of the generation, respectively. $D_{NO,air}$ is the molecular diffusion coefficient of NO in air. $S_{air,i}$ and $S_{alv,i}$ are the bronchial and alveolar lateral surfaces of the $i^{th}$ generation, respectively. $J_{air}$ and $J_{alv}$ are the NO exchange flux densities from airways (Eq.6) and alveoli (Eq.7), respectively. The model is radially expansible during a respiratory cycle. Thus $Q$, $\Omega'$, $\Omega$, $S_{air}$, $S_{alv}$, $J_{air}$ and $J_{alv}$ are time dependent.

The hypotheses on which Eq.8 relies may be found in (31). Solving Eq.8 provides NO concentration in any lung generation and for any respiratory phase.

### d) Airway caliber change

This layered model is a dynamic model, which can be used to simulate an airway caliber change in any bronchial generation. The outer radius of an airway is $R_\mu + \delta_\mu + \delta_E + \delta_M$, $R_\mu$ being the inner radius (see Fig.1C). During a bronchoconstriction, the surrounding muscle is contracting. Its shortening leads to a decrease of the outer radius, thus of the radii of each layers. Due to the volume conservation in each layer, their thicknesses ($\delta_\mu$, $\delta_E$ and $\delta_M$) increase, reducing the inner radius $R_\mu$ to a larger extent than the outer radius.

A constriction coefficient $\beta$ was defined, for the $i^{th}$ generation, as

$$\beta_i = 1 - \frac{R^2_{\mu,i}}{R^2_{\mu,i,0}} \qquad (9)$$

The subscript 0 indicates that the radius is evaluated with no bronchoconstriction occurring. Thus, $\beta$ compares the effective occlusion of the lumen before and after constriction, $\beta = 0$ indicating no constriction, and $\beta = 1$ indicating a total occlusion of the bronchus.

Bronchoconstrictions were simulated by comparing baseline ($\beta=0$) to obstructed state (with a given $\beta$) and bronchodilations were simulated by comparing an obstructed state (with a given $\beta$) to a non-obstructed state ($\beta=0$). In this way, $0<\beta<1$ may be used to characterize the two situations.

### e) Numerical simulations

The NO diffusion-convection equation (Eq.8) is solved numerically in a dimensionless form, in order to accelerate and stabilize the computations. An extensive description of the equations solving and an access to the associated Mathematica® codes can be found in Karamaoun *et al.* (16).

The simulations were performed with a 2 s inspiratory time of at a constant flow of 500 ml.s$^{-1}$ directly followed by a 20 s expiration at a constant flow of 50 ml.s$^{-1}$. The values and dimensions of the parameters common to all generations are presented in Table 1.

The following outcomes were considered:

- The total flux of NO in the $i^{th}$ generation: $JawNO_i = J_{air,i}.S_{air,i}$ (see Eq.6 and 8).
- The exhaled nitric oxide concentration, $F_ENO$, measured for an expiratory flow equal to 50 ml.s$^{-1}$. $F_ENO$ was defined as the NO concentration at the onset of the model (generation 0) at the end of expiration. The NO production per epithelial volume $Pr$ and consumption coefficient $k$ were adjusted in order to yield $F_ENO$ equal to 15 ppb in baseline conditions (*i.e.* β = 0 and without mucus layer)

The expiratory flow for $F_ENO$ computation is in line with the guidelines (2). The latter also recommend an inspiration up to total lung capacity (instead of 1 liter) and $F_ENO$ computed as the mean expiratory plateau measured for at least 3 s during an expiration of at least 6 s (instead of a single value after 20 s expiration). The impact of the deviations from these recommendations on simulated $F_ENO$ was evaluated in (31) and shown to be trivial.

## RESULTS

Table 2 summarizes bronchoconstriction simulations, giving the parameter considered, whether or not mucus layer or bronchoconstriction was involved, the site on which mucus layer or bronchoconstriction is acting, and the associated figure.

1) *Impacts of bronchoconstriction and mucus layer on total NO flux (JawNO)*

Figure 2 shows how a moderate ($\beta=0.5$) and a marked ($\beta=0.9$) constriction are acting on a bronchiole of generation 16, without (Panel A) or with (Panel C) a mucus layer. The red arrows allow appreciating JawNO reduction between the epithelium and the airway lumen. Panels B and D show JawNO as a function of the generation number in baseline and constricted situations with (Panel D) and without (Panel B) a 15 µm thick mucus layer coating the epithelium from generations 0 to 18.

2) *Impacts of bronchoconstriction and mucus layer on $F_ENO$*

Figure 3 evaluates the distinct effects of bronchoconstriction and mucus layer on $F_ENO$. Panel A shows the impact of bronchoconstriction (moderate and marked) progressively penetrating the bronchial tree free from mucus. Panel B presents the effect of a mucus layer on $F_ENO$ up to a given generation without muscle contraction and for different mucus thicknesses.

3) *Mucus layer and bronchoconstriction interaction effect on $F_ENO$*

Figure 4A presents the impact of a marked bronchoconstriction ($\beta=0.9$) progressively penetrating the bronchial tree whereas a mucus layer is coating generations 0 to 18. Different mucus thicknesses (including no mucus) were considered. Figure 4B shows the effect on $F_ENO$ of a marked bronchoconstriction ($\beta=0.9$) from generations 0 to 18 whereas a mucus layer is progressively coating the bronchial tree.

4) *Distinct response patterns to bronchodilation:*

Figure 5 summarizes the effects of an acute bronchodilation from a baseline constricted state (given $\beta$) to a non-constricted state ($\beta=0$) as a function of mucus thickness (abscissa) and baseline bronchoconstriction amplitude ($\beta$, ordinate). The solid lines delineate the combinations of mucus thickness and $\beta$ yielding a given $F_ENO$ change (in % from baseline) after bronchodilation. These iso-change lines are interrupted at regular intervals by the considered value of $F_ENO$ change. By example, the lower iso-line on Panel D corresponds to a 5% $F_ENO$ decrease (-5) and the upper line to a 15% decrease (-15). Panels A, B, C differs by the area of the bronchial tree on which the bronchodilation applies: from generation 0 to 6

(central airways), 0 to 15 (up to terminal bronchioles), and 0 to 18 (up to intra-acinar airways), respectively, all with a mucus layer coating generation 0 to 18. Panel D is analogous to Panel C with a mucus layer coating only generation 0 to 14. Iso-lines are colored as a function of $F_E$NO behavior: blue for a non-significant $F_E$NO change, green for $F_E$NO increase of at least 10% and red for $F_E$NO decrease of at least 10% after bronchodilation.

# DISCUSSION

The simulations presented in this work theoretically support the innovative concept that $F_ENO$ may play a role as a functional marker of acute airway caliber change, not implying change in inflammatory status. Indeed, the present work showed that caliber change may account for acute $F_ENO$ changes evidenced by experimental findings (7, 12, 13, 20, 30).

The first models of NO production and transport, notably accounting for the acute $F_ENO$ dependence on expiratory flow (28), were two-compartment models, *i.e.* conducting airways and alveolar zone, which considered only NO transport by convection (14, 15, 22, 29). Further models (26, 31) introduced axial diffusion transport and more realistic boundary conditions based on the Weibel's symmetrical morphometrical model (33). The present model (16) is the combination of the axial NO transport model (31) in the airway lumen and of the NO diffusion inside the airway tissues surrounding the lumen, the last feature being inspired by the Tsoukias *et al.* epithelial model (29). NO produced in the epithelium is either consumed (3), or is transported by diffusion to the blood surrounding the airways (considered as an infinite sink), or to the airway lumen where it is excreted through the epithelial surface. So, instead of being imposed as in previous models, the NO flux from the epithelium into the lumen (currently called JawNO), that enriches alveolar air during expiration, is depending on NO production and consumption inside the epithelium tissue and on the physical and geometrical properties of the tissue/air interface. Moreover, the present model allows to evaluate the effect of an airway caliber change on JawNO and $F_ENO$ by simulating the shortening or lengthening (contraction and relaxation) of the muscle layer surrounding the airway which, in turn, impacts the thickness of the epithelial layer and its interface area with the airway lumen. A previous model tackled the issue of bronchoconstriction effect on $F_ENO$ (32). However, it was dedicated to axial NO transport and only allowed to present different scenarios, resulting from somehow arbitrary assumptions about how JawNO might be affected by bronchoconstriction.

The present model shows that JawNO is mechanically, *i.e.* without NO production change, reduced by the decrease of the available epithelial surface through which NO is diffusing, this reduction being linked to the bronchoconstriction amplitude. Noteworthy, this effect becomes significant only in the small airways where the greater amount of production is concentrated.

The present model also showed that a mucus layer, not considered in previous models, dramatically affects JawNO in the small airways.

Up to the pre-acinar small airways, *i.e.* generations 15-16, $F_E NO$ decreases as a function of the bronchoconstriction amplitude and location, the deeper acts the bronchoconstriction, the greater is the $F_E NO$ decrease. This was experimentally observed with airway challenges using different provocative agents acting on different sites of the peripheral airways (19, 30), with comparable degrees of bronchoconstriction in the same subjects.
When constriction penetrates deeper, the axial molecular diffusion in peripheral airways begins to play a crucial role as already emphasized by previous models (26, 31). Indeed, during expiration, the NO concentration gradient between pre-acinar bronchioles and the alveolar compartment induces a NO diffusion flux, the so-called "back-diffusion", towards alveoli through intra-acinar airways, removing NO molecules from the expiration flow. Acinar airway constriction impairs the back-diffusion, by reducing the surface through which diffusion occurs, and allows more NO molecules to be expired. Consequently, $F_E NO$ increases despite an overall JawNO decrease.

Besides airway caliber change, a unique feature of the present model is the opportunity to consider a mucus layer (assimilated to water) coating the airway epithelium. Indeed, an increased and more peripheral mucus layer with goblet cell hyperplasia has been evidenced in chronic asthma, even mild, compared to healthy subjects (1, 9). This layer increases the length of the diffusion pathway from epithelium to the lumen, resulting in a reduction of both JawNO and $F_E NO$, even when it occurred in the acinus (up to generation 18). Indeed, even a relatively thin layer may impair NO diffusion from the epithelium into the lumen but will not greatly affect the axial back diffusion flux. Compared to a situation without mucus and for a given airway caliber reduction up to pre-acinar airways, the presence of a mucus layer amplifies $F_E NO$ decrease by 200 to 300%. Moreover, if constriction goes deeper into the acinus, no more $F_E NO$ increase occurs. To allow $F_E NO$ increase due to intra-acinar constriction, the mucus layer has to be no more distal than the 14-15 generations.

Though back-diffusion was evidenced by heliox experiments in healthy subjects (17, 25) and in asthma patients (17), a $F_E NO$ increase due to intra-acinar caliber reduction has never been experimentally observed, likely because no currently used provocative agent has such a peripheral effect. Fortunately, the mirror manoeuver, *i.e.* an acute bronchodilation going from a baseline constricted state to a non- (or less) constricted state, was shown, by ventilation distribution tests, to act up to intra-acinar airways in one third of a cohort of asthma patients. These patients concomitantly exhibited a marked $F_E NO$ decrease. The two other thirds of asthma patients exhibited an increase or no change in $F_E NO$ (20). These observations constitute the opportunity to theoretically estimate whether combinations of mucus thickness and bronchodilation amplitude may mimic them and whether they may be informative about the location of the bronchodilation. This is the purpose of Figure 5 which presents, as contour plots, the combinations of mucus thickness and bronchodilation amplitude yielding a given $F_E NO$ change. These combinations were simulated for a central dilation (Fig.5A, generations 0 to 6), a dilation up to pre-acinar bronchioles (Fig.5B, generations 0 to 15), and up to intra-acinar airways (Fig.5C, generations 0 to 18). In panels A to C, dilation is associated with mucus coating generations 0 to 18 and, panel D, with mucus coating only generations 0 to 14. Though some very specific β/mucus thickness combinations may lead to a non-significant $F_E NO$ change after peripheral bronchodilation (Panel B and C), it came out that this feature is likely associated with dilation of central airways (blue lines in Fig5A). Conversely, a substantial $F_E NO$ change, whatever the direction of the change, rules out a dilation limited to central airways. $F_E NO$ decrease (red lines) is associated only with intra-acinar airways dilation (Fig.5D and 5C), even moderate, which boosts the back-diffusion and, thus, removes more NO molecules from the expiratory flow. It is to be noted that this effect is not observed if a substantial mucus layer goes further than the 14$^{th}$ generation. $F_E NO$ increase (green lines) is also always linked to peripheral airways. It may be essentially observed with marked dilations in pre-acinar airways (Fig.5B), for a large range of mucus thicknesses, but also in intra-acinar airways if a thick mucus layer is present. Altogether, the present simulations confirmed the link between distinct $F_E NO$ behaviors that were experimentally evidenced after bronchodilation, and specific sites of actions (20), which opens the way to a very simple and informative test, *i.e.* NO measurement before and after acute bronchodilation. Moreover, they allowed making assumptions about the presence of a mucus layer in very peripheral airways, relevant issue (23) out of the reach of classical lung function tests.

Like all model studies, this work has limitations essentially consisting in over-simplifications. We considered a constant epithelial thickness along the bronchial tree and a constant NO production per unit epithelial volume. This may have led to overestimate the NO production in the very peripheral airways. However, in healthy subjects, Boers *et al.* (5) found an increasing number of Clara cells in terminal to respiratory bronchioles and Shaul *et al.* (24) showed that endothelial NO synthase (eNOS) is expressed in cultured cells of the same lineage than Clara cells (NCI-H441 human bronchiolar epithelial cells). This suggests an increase of NO production in peripheral airways that is likely amplified in asthma patients (30). Finally, the distribution of NO production assumed in the present simulations is very close to that deduced from experimental data (18, 30). We also considered a constant mucus layer thickness in central as well as in peripheral airways. However, it appeared that the mucus layer in the central airways barely affected JawNO or $F_E$NO, whatever its thickness. In peripheral airways we considered mucus thickness up to 15μm which may seem overestimated. Indeed, a 15 μm thick layer would represent ±10% of a $16^{th}$ generation bronchiole lumen area where ±5% is typically described in asthma patients (1). However, we assimilated mucus to water, which certainly overestimated the NO diffusivity through the mucus. The decrease of NO diffusivity with increased mucus density, not yet established should be included in further refinements of the model.

Finally, it is to be noted that this new role of $F_E$NO as a marker of airway caliber change after an acute intervention is theoretically not limited to asthma. $F_E$NO, a biomarker detectable even in the absence of inflammation, may be used to detect peripheral airway caliber change in other pathological situations, such as chronic obstructive disease, even with active smoking, or cystic fibrosis, in which $F_E$NO is considered having no or little usefulness as an inflammatory marker. These certainly are opportunities for further developments.

In conclusion, the present simulations showed that the link between $F_E$NO behavior after an airway caliber change and the localization (and the amplitude) of this change is theoretically sound. This opens the way for a new area of research in which $F_E$NO, besides its undeniable utility as an inflammation marker, may play a new role as a peripheral functional marker.

# LEGENDS TO THE FIGURES

Figure 1:

Schematic representation of an airway: Panel A: transversal cross-section. Panel B: longitudinal cross-section. The white zone is the airway lumen; the grey zone is the airway wall. The $z$ axis is dedicated to the axial NO transport in the airway lumen. Panel C is a close-up of the rectangle of Panel A describing the structure of the airway wall. From outer to inner part: blood (where NO tissues concentration $C_t = 0$), muscle (thickness $\delta_M$, no NO production and consumption $-k.C_t$), epithelium (thickness $\delta_E$, NO production per unit volume $Pr$ and consumption $-k.C_t$), mucus (thickness $\delta_\mu$, no production and no consumption). $R_\mu$ is the radius of the unobstructed part of the airway lumen. The $x$ axis is dedicated to the NO transport in the airway wall. $J_{air}$ is the NO flux density excreted into the airway lumen.

Figure 2:

Panel A: cross-section of a generation 16 bronchiole without mucus layer, without constriction ($\beta=0$), with a moderate constriction ($\beta=0.5$), and with a marked constriction ($\beta=0.9$) from generations 0 to 18. Red arrows are proportional to JawNO. Panel B: JawNO as a function of the generation number for the constriction depicted on Panel A. Panels C and D: analogous to Panels A and B, with a mucus layer of 15 µm thickness covering generations 0 to 18.

Figure 3:

Panel A: $F_ENO$ change (in % of baseline) when a moderate (grey line) or a marked (black line) bronchoconstriction penetrates the bronchial tree (from generation 0 up to …) free from mucus. The baseline situation is $\beta=0$. Panel B: $F_ENO$ change (in % of baseline) when a mucus layer of 5 (light grey), 10 (dark grey) and 15 (black) µm thickness progressively coats the bronchial tree (from generation 0 up to …) without muscle length change ($\beta=0$). The baseline situation is no mucus layer.

Figure 4:

Panel A: $F_ENO$ change (in % of baseline) due to a marked bronchoconstriction ($\beta=0.9$) penetrating the bronchial tree (from generation 0 up to …) either without mucus (dashed line) or with a mucus layer of 5 (light grey), 10 (dark grey) and 15 (black) µm thickness coating the generations 0 to 18. Baseline situation is $\beta=0$ and mucus layer, if any. Panel B: $F_ENO$ change (in % of baseline) due to a marked bronchoconstriction ($\beta=0.9$) from generations 0 to 18 when a mucus layer of 5 (light grey), 10 (dark grey) and 15 (black) µm thickness progressively coats the bronchial tree (from generation 0 up to …). Baseline situation is $\beta=0$ and mucus layer up to the given generation.

Figure 5:

Effect of an acute bronchodilation on $F_E$NO from a baseline constricted state to a non-constricted state. Panels A-D: contour plots of iso-$\Delta F_E$NO% as a function of mucus thickness and baseline bronchoconstriction amplitude ($\beta$). Panels A, B, C correspond to a baseline bronchoconstriction from generation 0 to 6, 0 to 15, and 0 to 18, respectively, with a mucus layer coating generation 0 to 18. Panel D is analogous to Panel C with a mucus layer coating generation 0 to 14. Color lines delineate the mucus thickness/$\beta$ combinations corresponding to a non-significant $F_E$NO change (blue lines), a $F_E$NO increase (green lines) and a $F_E$NO decrease (red lines) after bronchodilation.

Table 1

| Parameter | Description | Value | Units | References |
|---|---|---|---|---|
| $\delta_{E,0}$ | Airway epithelium thickness | 0.0015 | cm | (10) |
| $\delta_{M,0}$ | Muscle thickness | 0.0030 | cm | (10) |
| $D_{NO,air}$ | NO diffusion coefficient in air | 0.217 | $cm^2.s^{-1}$ | (31) |
| $D_{NO,t}$ | NO diffusion coefficient in tissues | $3.3\ 10^{-5}$ | $cm^2.s^{-1}$ | (29) |
| $\gamma$ | Correction factor | $2.545\ 10^4$ | $cm^3.mol^{-1}$ | (16) |
| k | Kinetic constant of NO consumption in the tissues | 2.001 | $s^{-1}$ | Adapted from (29) |
| $\lambda_{t:air}$ | Tissue/air equilibrium constant | $1.64\ 10^{-6}$ | $molNO.cm^{-3}$ | Adapted from (29) |
| Pr | NO epithelial production per unit tissue volume | $5.17\ 10^{-12}$ | $molNO.cm^{-3}.s^{-1}$ | Adapted from (29) |
| $P_{alv}$ | Total NO alveolar production | $3.167\ 10^{-6}$ | $mlNO.s^{-1}$ | (22) |
| $U_{alv}$ | NO consumption in alveolar tissues | 1558 | $cm^3.s^{-1}$ | (22) |

Table 2: Bronchoconstriction simulations summary

| Figure | Outcome | Mucus thickness | Localization[*] | Bronchoconstriction | Localization[*] |
|---|---|---|---|---|---|
| Fig2A/2B | JawNO | no | - | $\beta=0.5$; $\beta=0.9$ | 0 - 18 |
| Fig2C/2D | JawNO | 5, 10, 15 µm | 0 - 18 | $\beta=0.5$; $\beta=0.9$ | 0 - 18 |
| Fig3A | $F_E NO$ | no | - | $\beta=0.5$; $\beta=0.9$ | $0 - n^{**}$ |
| Fig3B | $F_E NO$ | 5, 10, 15 µm | $0 - n^{**}$ | no | - |
| Fig4A | $F_E NO$ | 5, 10, 15 µm | 0 - 18 | $\beta=0.9$ | 0 - 18 |
| Fig4B | $F_E NO$ | 5, 10, 15 µm | $0 - n^{**}$ | $\beta=0.9$ | 0 - 18 |

[*]: generation number where mucus layer and/or bronchoconstriction are/is present; [**]: from generation 0 up to a given generation n.

# REFERENCES

1.  **Aikawa T, Shimura S, Sasaki H, Ebina M, and Takishima T**. Marked goblet cell hyperplasia with mucus accumulation in the airways of patients who died of severe acute asthma attack. *Chest* 101: 916-921, 1992.
2.  **American Thoracic S, and European Respiratory S**. ATS/ERS recommendations for standardized procedures for the online and offline measurement of exhaled lower respiratory nitric oxide and nasal nitric oxide, 2005. *Am J Respir Crit Care Med* 171: 912-930, 2005.
3.  **Beckman JS, and Koppenol WH**. Nitric oxide, superoxide, and peroxynitrite: the good, the bad, and ugly. *Am J Physiol* 271: C1424-1437, 1996.
4.  **Berry MA, Shaw DE, Green RH, Brightling CE, Wardlaw AJ, and Pavord ID**. The use of exhaled nitric oxide concentration to identify eosinophilic airway inflammation: an observational study in adults with asthma. *Clin Exp Allergy* 35: 1175-1179, 2005.
5.  **Boers JE, Ambergen AW, and Thunnissen FB**. Number and proliferation of clara cells in normal human airway epithelium. *Am J Respir Crit Care Med* 159: 1585-1591, 1999.
6.  **Borland CD, and Higenbottam TW**. A simultaneous single breath measurement of pulmonary diffusing capacity with nitric oxide and carbon monoxide. *Eur Respir J* 2: 56-63, 1989.
7.  **de Gouw HW, Hendriks J, Woltman AM, Twiss IM, and Sterk PJ**. Exhaled nitric oxide (NO) is reduced shortly after bronchoconstriction to direct and indirect stimuli in asthma. *Am J Respir Crit Care Med* 158: 315-319, 1998.
8.  **Dweik RA, Boggs PB, Erzurum SC, Irvin CG, Leigh MW, Lundberg JO, Olin AC, Plummer AL, Taylor DR, and American Thoracic Society Committee on Interpretation of Exhaled Nitric Oxide Levels for Clinical A**. An official ATS clinical practice guideline: interpretation of exhaled nitric oxide levels (FENO) for clinical applications. *Am J Respir Crit Care Med* 184: 602-615, 2011.
9.  **Fahy JV, and Dickey BF**. Airway mucus function and dysfunction. *N Engl J Med* 363: 2233-2247, 2010.
10. **Farmer SG, and Hay DWP**. *The Airway Epithelium: Physiology, Pathophysiology and Pharmacoloy*. New York, NY: Marcel Dekker, 1991.
11. **Guo FH, De Raeve HR, Rice TW, Stuehr DJ, Thunnissen FB, and Erzurum SC**. Continuous nitric oxide synthesis by inducible nitric oxide synthase in normal human airway epithelium in vivo. *Proc Natl Acad Sci U S A* 92: 7809-7813, 1995.
12. **Haccuria A, Michils A, Michiels S, and Van Muylem A**. Exhaled nitric oxide: a biomarker integrating both lung function and airway inflammation changes. *J Allergy Clin Immunol* 134: 554-559, 2014.
13. **Ho LP, Wood FT, Robson A, Innes JA, and Greening AP**. The current single exhalation method of measuring exhales nitric oxide is affected by airway calibre. *Eur Respir J* 15: 1009-1013, 2000.
14. **Hogman M, Holmkvist T, Wegener T, Emtner M, Andersson M, Hedenstrom H, and Merilainen P**. Extended NO analysis applied to patients with COPD, allergic asthma and allergic rhinitis. *Respir Med* 96: 24-30, 2002.
15. **Jorres RA**. Modelling the production of nitric oxide within the human airways. *Eur Respir J* 16: 555-560, 2000.
16. **Karamaoun C, Van Muylem A, and Haut B**. Modeling of the Nitric Oxide Transport in the Human Lungs. *Front Physiol* 7: 255, 2016.
17. **Kerckx Y, Michils A, and Van Muylem A**. Airway contribution to alveolar nitric oxide in healthy subjects and stable asthma patients. *J Appl Physiol (1985)* 104: 918-924, 2008.
18. **Kerckx Y, and Van Muylem A**. Axial distribution heterogeneity of nitric oxide airway production in healthy adults. *J Appl Physiol (1985)* 106: 1832-1839, 2009.
19. **Michils A, Elkrim Y, Haccuria A, and Van Muylem A**. Adenosine 5'-monophosphate challenge elicits a more peripheral airway response than methacholine challenge. *J Appl Physiol (1985)* 110: 1241-1247, 2011.

20. **Michils A, Malinovschi A, Haccuria A, Michiels S, and Van Muylem A**. Different patterns of exhaled nitric oxide response to beta2-agonists in asthmatic patients according to the site of bronchodilation. *J Allergy Clin Immunol* 137: 806-812, 2016.
21. **Piacentini GL, Bodini A, Peroni DG, Miraglia del Giudice M, Jr., Costella S, and Boner AL**. Reduction in exhaled nitric oxide immediately after methacholine challenge in asthmatic children. *Thorax* 57: 771-773, 2002.
22. **Pietropaoli AP, Perillo IB, Torres A, Perkins PT, Frasier LM, Utell MJ, Frampton MW, and Hyde RW**. Simultaneous measurement of nitric oxide production by conducting and alveolar airways of humans. *J Appl Physiol (1985)* 87: 1532-1542, 1999.
23. **Rogers DF**. Airway mucus hypersecretion in asthma: an undervalued pathology? *Curr Opin Pharmacol* 4: 241-250, 2004.
24. **Shaul PW, North AJ, Wu LC, Wells LB, Brannon TS, Lau KS, Michel T, Margraf LR, and Star RA**. Endothelial nitric oxide synthase is expressed in cultured human bronchiolar epithelium. *J Clin Invest* 94: 2231-2236, 1994.
25. **Shin HW, Condorelli P, and George SC**. Examining axial diffusion of nitric oxide in the lungs using heliox and breath hold. *J Appl Physiol (1985)* 100: 623-630, 2006.
26. **Shin HW, and George SC**. Impact of axial diffusion on nitric oxide exchange in the lungs. *J Appl Physiol (1985)* 93: 2070-2080, 2002.
27. **Silkoff PE, McClean PA, Caramori M, Slutsky AS, and Zamel N**. A significant proportion of exhaled nitric oxide arises in large airways in normal subjects. *Respir Physiol* 113: 33-38, 1998.
28. **Silkoff PE, McClean PA, Slutsky AS, Furlott HG, Hoffstein E, Wakita S, Chapman KR, Szalai JP, and Zamel N**. Marked flow-dependence of exhaled nitric oxide using a new technique to exclude nasal nitric oxide. *Am J Respir Crit Care Med* 155: 260-267, 1997.
29. **Tsoukias NM, and George SC**. A two-compartment model of pulmonary nitric oxide exchange dynamics. *J Appl Physiol (1985)* 85: 653-666, 1998.
30. **Van Muylem A, Kerckx Y, and Michils A**. Axial distribution of nitric oxide airway production in asthma patients. *Respir Physiol Neurobiol* 185: 313-318, 2013.
31. **Van Muylem A, Noel C, and Paiva M**. Modeling of impact of gas molecular diffusion on nitric oxide expired profile. *J Appl Physiol (1985)* 94: 119-127, 2003.
32. **Verbanck S, Kerckx Y, Schuermans D, Vincken W, Paiva M, and Van Muylem A**. Effect of airways constriction on exhaled nitric oxide. *J Appl Physiol (1985)* 104: 925-930, 2008.
33. **Weibel ER**. *Morphometry of the Human Lung*. New-York: Academic, 1963.